\documentclass[twocolumn,aps,prb,showpacs]{revtex4}
\usepackage{graphicx}
\usepackage{epsf}

\begin{document}

\title{Electronic Structure of Hyperkagome Na$_4$Ir$_3$O$_8$}

\author{M. R. Norman and T. Micklitz}
\affiliation{Materials Science Division, Argonne National Laboratory, Argonne, IL 60439}

\begin{abstract}
We investigate the electronic structure of the frustrated magnet Na$_4$Ir$_3$O$_8$
using density functional theory.  Due to strong spin-orbit coupling, the hyperkagome
lattice is characterized by a half-filled complex of $d$ states,
making it a cubic iridium analogue of the
high temperature superconducting cuprates.  The implications of our results 
for this unique material are discussed.
\end{abstract}
\date{\today}
\maketitle

\section{Introduction}

Recently, several materials have been identified \cite{leeNV} as possible candidates for a
long sought state of matter - the quantum spin liquid.\cite{phil73}
One of these candidates, Na$_4$Ir$_3$O$_8$, is characterized by a three dimensional cubic space
group, where the iridium sites form a hyperkagome lattice of corner sharing triangles.
This remarkable material is an insulator with a large Curie-Weiss temperature (650 K) and
 a large effective moment (1.96 $\mu_B$), yet does not exhibit any sign of magnetic order down to the
 lowest measured temperature.\cite{okamoto}
 
 There have been a number of theoretical studies of this iridate \cite{lawler} which
 attempt to address the nature of the spin liquid ground state.  These studies invariably assume
 that the material is described by an isotropic Heisenberg model, since anisotropy acts to
 stabilize magnetic order.  Isotropy would be a surprise
given the strong spin-orbit coupling for the iridium ions.  But in an exhaustive
 study, Chen and Balents \cite{balents} formulated under what conditions such an effective model 
 might be realized.  To address various issues raised in that paper, we calculated the electronic 
 structure of this material within the local density approximation, then performed a tight binding fit
 in order to estimate the various exchange integrals which enter the spin Hamiltonian for this
 frustrated magnet.  We find that the spin Hamiltonian should be strongly anisotropic,
 but offer two possible scenarios where isotropy might be restored.
 
 \section{Calculational Details}
 
 Although Na$_4$Ir$_3$O$_8$ has a cubic space group, the unit cell is
 quite complicated, comprising four formula units (60 atoms).\cite{okamoto}  Moreover, the lattice
 breaks inversion symmetry.  The space group, P4$_1$32, is also found in the
 non-centrosymmetric superconductors Li$_2$Pd$_3$B and Li$_2$Pt$_3$B.\cite{pickett}
 The material is formed from distorted IrO$_6$ octahedra, each comprised of two different oxygen
 sites - four of type 2 (O2), two of type 1 (O1).  The latter should not be thought of as `apical' oxygens
 in the cuprate sense, since they are not related by inversion through the iridium site.
 Rather, each octahedron is described by a C$_2$ axis along a (110) direction.\cite{balents}
 There are twelve
 of these axes, one associated with each of the twelve iridium ions in the unit cell.
 Each oxygen ion is also at the center of a distorted octahedron, with O1
 octahedra
 of the form OIr$_3$Na$_3$, and O2 octahedra of the form OIr$_2$Na$_4$.  Each oxygen ion
 is also surrounded by a very distorted array of twelve other oxygen ions along
 approximate (110) directions.
 Finally, each iridium ion is surrounded by four other iridium ions,
 also along approximate (110) directions.  These form the corner sharing triangles that characterize the
 hyperkagome lattice.  A list of the first few coordination shells around an iridium ion is given in Table I.
 
\begin{table}
\caption{Coordination shells around an iridium ion in the Na$_4$Ir$_3$O$_8$ structure out
to a distance of 0.49$a$.
Listed is the type of ion, their number, (approximate) direction, 
and distance (in units of the lattice constant, $a$).}
\begin{ruledtabular}
\begin{tabular}{lccc}
type & number & direction & distance \\
\colrule
O2 &  2  & 100 & 0.2273 \\
O2 &  2  & 100 & 0.2280 \\
O1 &  2  & 100 & 0.2333 \\
Ir & 4  & 110 & 0.3464 \\
Na3 &  2  & 110 & 0.3501 \\
Na3 &  2  & 110 & 0.3587 \\
Na1 &  2  & 110 & 0.3612 \\
Na3 & 1  & 110 & 0.3627  \\
Na2 &  1  & 110 & 0.3683 \\
O2 &  2  & 111 & 0.4065 \\
O2 &  2  & 111 & 0.4225 \\
O1 & 2  & 111 & 0.4375 \\
O2 &  2  & 111 & 0.4838 \\
\end{tabular}
\end{ruledtabular}
\end{table}

 When considering an electronic structure calculation for this material,
 the first issue is that there are three
 different types of sodium ions, two of which are partially occupied with a stochiometry 
 of 75\%.\cite{okamoto}  Therefore, to form a simple unit cell, some approximation has to
 be made for these other two sodium sites.  We choose to fully occupy the $12d$ sites (sodium
 type 3)  and replace the $4a$ sites (sodium type 2) by empty spheres.  The resulting lattice
 has 64 sites in the unit cell - 12 Ir, 8 O1, 24 O2, 4 Na1, 4 empty (i.e., Na2), and 12 Na3.
 
The calculations in this paper were performed using the local density approximation (LDA)
within a linear muffin tin orbital scheme.\cite{LMTO}  Overlap (combined) corrections
were incorporated, but the Coulomb potential at each site was treated in a spherical approximation.
The exchange-correlation potential used was that of Hedin and Lundqvist.\cite{HL}
Noting that the cubic lattice constant is 8.985$\AA$, a muffin tin radius of 1.238$\AA$ was used
on the iridium sites, and 1.425$\AA$ was used on all other sites.
These radii were chosen based on the various near neighbor separations.
$s,p$ basis functions were used on all sites (including the empty ones) except for the
iridium sites, where $d$ basis functions were also included.  For the scalar relativistic calculation
(all relativistic effects but spin-orbit),
this resulted in a secular matrix of dimension 316, which was double this (632) for the full
relativistic (Dirac equation) calculation.
The calculations were converged using 56 $k$ points in the irreducible wedge
of the Brillouin zone.
We should note that the Brillouin zone for this material is simple cubic.  Though inversion
symmetry is broken, one can restrict the irreducible wedge to 1/48th of the zone.
The eigenvalues at a general $k$ point are doubly (Kramers) degenerate unless spin-orbit is
incorporated, where this degeneracy is broken.
After convergence, 165 $k$ points were generated in the irreducible wedge.  The
eigenvalues were then fit using a Fourier spline series \cite{SPLINE} with 560 lattice
periodic functions.  A linear tetrahedron method \cite{LT} was used to
evaluate the density of states,
with the zone broken down into 48 $\cdot$ 8$^5$ tetrahedra.

\section{Results}

\begin{figure}
\centerline{
\includegraphics[width=3.4in]{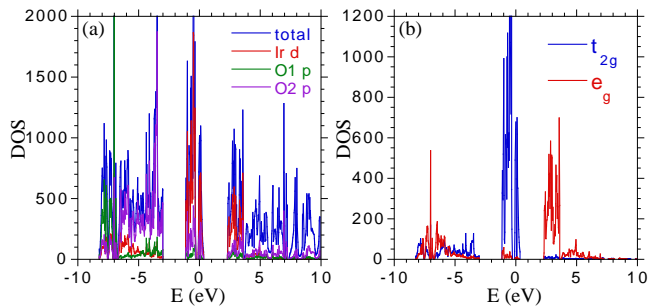}
}
\caption{(Color online) Density of states (DOS) in states/Ryd
for the scalar relativistic calculation of Na$_4$Ir$_3$O$_8$.
The Fermi energy, $E_F$, is at zero.}
\label{fig1}
\end{figure}

\begin{figure}
\centerline{
\includegraphics[width=3.4in]{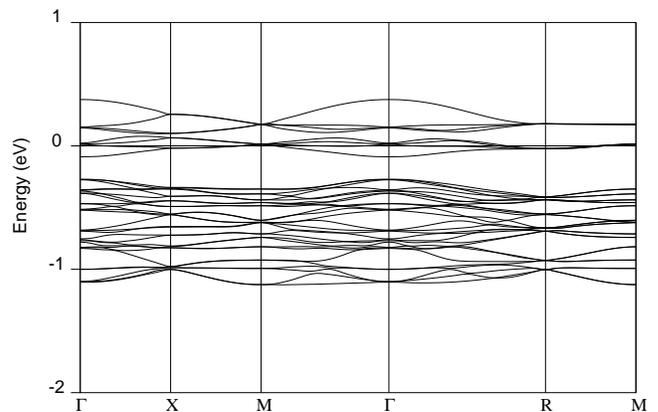}
}
\caption{Energy bands of the t$_{2g}$ complex for the scalar relativistic calculation of Na$_4$Ir$_3$O$_8$.  $\Gamma=(0,0,0)$, $X=(1,0,0)$, $M=(1,1,0)$ and $R=(1,1,1)$ in units of $\pi/a$.
The horizontal line marks $E_F$.}
\label{fig2}
\end{figure}

\begin{figure}
\centerline{
\includegraphics[width=3.4in]{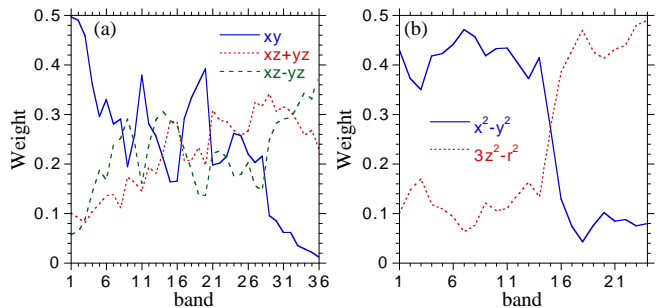}
}
\caption{(Color online)
Decomposition of the (a) t$_{2g}$ states and (b) e$_g$ states, averaged over each band,
into their individual components relative to the C$_2$ axis of the distorted IrO$_6$ octahedra
(scalar relativistic calculation).}
\label{fig3}
\end{figure}

We first present results from the scalar relativistic calculation, with the density of states shown in Fig.~1.
The O1 states form a narrow complex of bands ranging
from -8.2 to -6.7 eV, followed by the wider O2 complex ranging from -6.7 to -3 eV 
(there is only a narrow gap between these two complexes).  There is a gap
of 1.9 eV to the t$_{2g}$ complex of iridium d bands ranging  from -1.1 eV to 0.4 eV.
This is followed by a crystal field gap of 1.9 eV to an e$_g$ complex ranging from 2.3 eV
to 3.6 eV.  For energies above this, the states are primarily from the sodium sites.
The main complex of interest is the t$_{2g}$ one (Fig.~2).  It is split into an array of 28 bands (each
Kramers degenerate)
separated by a gap of 0.2 eV to an upper array of 8 bands.  This basic splitting of 28 and 8
arises from the t$_{pd}$ hopping integral, as identified from tight binding fits, and is related
to the four-fold degeneracy of the eigenvalues at the $R$ point of the zone.
There is a crystal field splitting of the $t_{2g}$ states which was identified by projecting
onto the three crystal field components due to the C$_2$ axis,
with the $xy$ states (for a 110 axis) lying lowest in energy, but it is rather ill defined (Fig.~3) and is small
relative to the overall $t_{2g}$ bandwidth.
The t$_{2g}$ complex 
is composed of 70\% Ir 5d character, 20\% O2 2p character, and 4\% O1 2p character.  The centroid
of the e$_g$ complex relative to the t$_{2g}$ one is 3.5 eV, with the t$_{2g}$-e$_g$ admixture
(due to the low iridium site symmetry) quite small.
The density of states at the Fermi energy (per Ir per spin) is 3.35 states/eV.  
The Fermi surface (not shown) is composed of small pockets, two about the $R$ point (electron-like),
two about the $M$ point (hole-like), and one around the $\Gamma$ point (hole-like).

\begin{figure}
\centerline{
\includegraphics[width=3.4in]{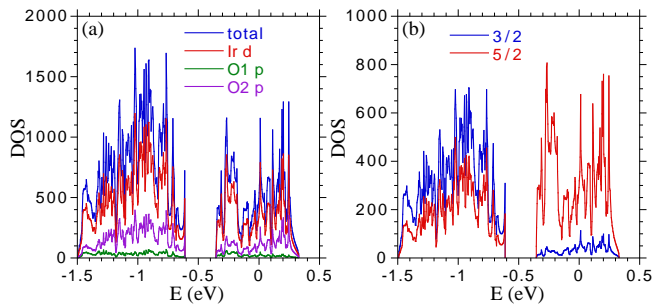}
}
\caption{(Color online) Density of states (DOS) in states/Ryd
for the full relativistic calculation of Na$_4$Ir$_3$O$_8$.}
\label{fig4}
\end{figure}

\begin{figure}
\centerline{
\includegraphics[width=3.4in]{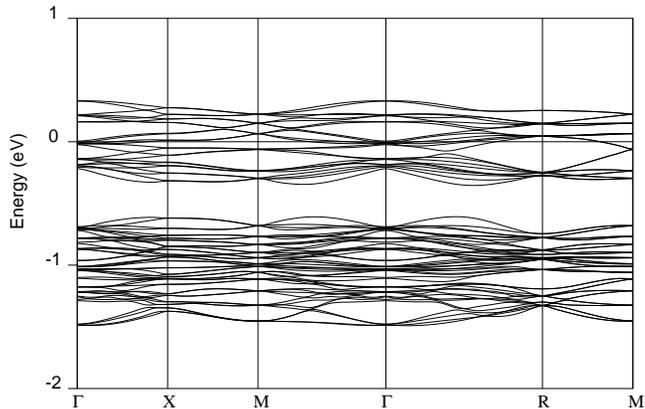}
}
\caption{Energy bands of the t$_{2g}$ complex for the full relativistic calculation of Na$_4$Ir$_3$O$_8$. The horizontal line marks $E_F$.}
\label{fig5}
\end{figure}

\begin{figure}
\centerline{
\includegraphics[width=2.4in]{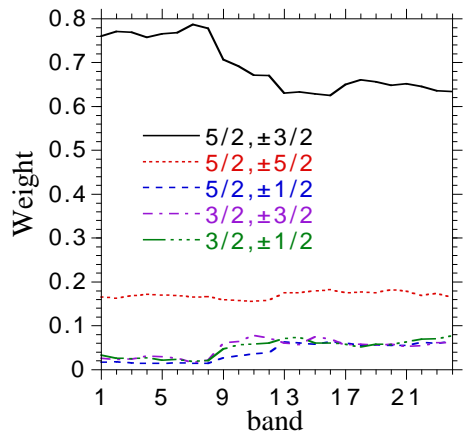}
}
\caption{(Color online)
Decomposition of the states of the upper t$_{2g}$ `doublet' complex, averaged over 
each band,
into their atomic spin-orbit components, $j, \pm m_j$ (full relativistic calculation).
The weights are normalized to the total d weight.}
\label{fig6}
\end{figure}

\begin{figure}
\centerline{
\includegraphics[width=3.4in]{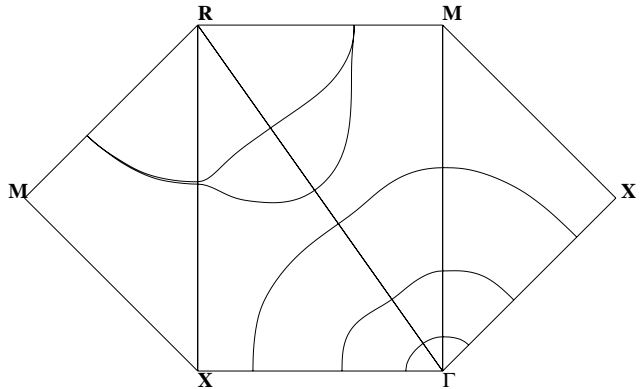}
}
\caption{Fermi surface for the full relativistic calculation of Na$_4$Ir$_3$O$_8$
in the symmetry planes of the zone.}
\label{fig7}
\end{figure}

Spin-orbit coupling has a profound impact on these results, as can be seen in the remaining
figures.  The t$_{2g}$ states now split (Figs.~4 and 5) into a lower complex of 48 bands and an upper complex
of 24 bands (the number of bands are now doubled because of the coupling of inversion breaking
with spin-orbit), with the e$_g$ complex composed of another 48 bands.
The  upper t$_{2g}$ complex is half-filled,
with the Ir d states mostly of $j=5/2$ character, with weak admixture
of $j=3/2$ character (Fig.~4).  As explained by Chen and Balents,\cite{balents} and found
as well for the electronic structure of Sr$_2$IrO$_4$,\cite{jjyu} this behavior is due to crystal
field admixture of the $j=5/2$ and $j=3/2$ states, which acts to form a mixed lower (t$_{2g}$)
quartet of states
(i.e., four per iridium site), a mixed (e$_g$)
upper quartet, with an unmixed (t$_{2g}$, $j=5/2$) doublet in between that is half-filled.
This doublet is predicted
to have 83\% $j=5/2, m_j=\pm 3/2$ and 17\% $j=5/2, m_j=\pm 5/2$ character, which is in good
agreement with the results we find for the lowest 8 bands of this complex (Fig.~6).  The upper 16 bands
have a slightly reduced $m_j=\pm 3/2$ weight due to weak admixture with $m_j=\pm 1/2$ states as well
as $j=3/2$ states, which is a consequence of the low iridium site symmetry.
Again, the density of states at the Fermi energy is not large - per Ir per spin
it is 1.51 states/eV.  The Fermi surface is composed of three electron like sheets around
$\Gamma$, and two hole-like sheets around the $R$ point (Fig.~7).  The bands are
doubly degenerate on the zone face due to the non-symmorphic nature
of the space group.

We now wish to connect these results to the insulating behavior of the actual
material.  In the
case of Sr$_2$IrO$_4$, inclusion of the Coulomb repulsion
is not sufficient to open an energy gap unless spin-orbit is
included.  In essence, spin-orbit produces a half-filled doublet which can then `Mott-Hubbardize'
and form an insulator, similar to what occurs in the cuprates.\cite{phil87}  Only a small $U$ 
is needed to cause this.\cite{jjyu}  Similar considerations should apply to Na$_4$Ir$_3$O$_8$.
From the centroids of the two quartets and the doublet, we have inferred the parameters $\Delta_{CF}$
(splitting of t$_{2g}$ and e$_g$) and $\lambda$ (coefficient of $l \cdot s$ for the spin-orbit coupling)
of 3.43 eV and 0.58 eV, respectively.

\section{Tight Binding Analysis}

To investigate these issues further, we performed tight binding fits to the electronic structure, restricting
to Ir 5d and O 2p orbitals.
This involved generating Slater-Koster parameters,\cite{SK} and then performing a least squares fit
to the band eigenvalues using Powell's method.\cite{NR}  Because of the distorted
nature of the lattice, there are a large number of tight binding parameters, even if one restricts
to near neighbors.
This is because there is not a unique distance for several of the atom combinations
(Ir-O2, O1-O1, O1-O2, and O2-O2).
To compensate for this, we assumed that the tight binding
parameters for each atom combination vary as the fourth power of the distance.  This assumption works
well for the O-O hoppings in IrO$_2$ \cite{iro2} (note from Table I that the two Ir-O2 distances are 
almost identical).
This reduced us to four t$_{pd}$ parameters ($\sigma$ and $\pi$ for Ir-O1 and Ir-O2), three
t$_{dd}$ parameters ($\sigma$, $\pi$, and $\delta$ for Ir-Ir), six t$_{pp}$ parameters ($\sigma$ 
and $\pi$ for O1-O1, O1-O2, and O2-O2), four on-site energies
(2p energies for O1 and O2, and t$_{2g}$ and e$_g$ energies), and the spin-orbit coupling
of the Ir $d$ ($\lambda$), for a total of 18 parameters.  We have explored putting in the residual
crystal field splittings due to the low Ir site symmetry (Fig.~3), but this did not lead to improved fits.

We first converged the fit for the scalar relativistic calculation, using 42 eigenvalues
(top and bottom of the O1 complex, O2 complex, 
e$_g$ complex,  and all 36 t$_{2g}$ bands) at the four symmetry points
($\Gamma$, $X$, $M$, and $R$).  As input, we used IrO$_2$ hopping parameters \cite{iro2}
scaled to the sodium iridate distances, as well as on-site energies estimated from averages
extracted from the $l$ decomposed density of states.
The  fit (not shown) yielded a band structure in good agreement with Fig.~2.
We then used these fit parameters as input to fitting the full relativistic calculation,
with the same set of eigenvalues (now numbering 78) at each of the 4 $k$ points.
The fit gives a good description of the bands and 
Fermi surface, as can be seen in Figs.~8 and 9.  These fit values are listed in Table II.  One can see
that the t$_{pd}$ and t$_{dd}$ parameters have comparable magnitude, and that the O1 and O2
integrals are quite different.  Note that the difference of the t$_{2g}$-e$_g$ energies is only 2.60 eV.
The difference from the centroid splitting of 3.43 eV quoted earlier is due to the contribution of the 
hoppings to the splitting.

\begin{figure}
\centerline{
\includegraphics[width=3.4in]{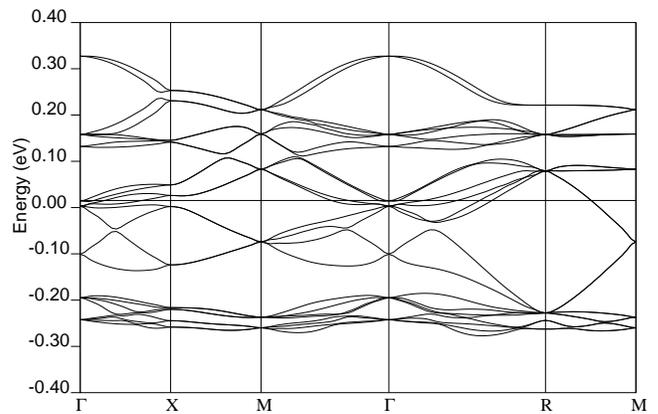}
}
\caption{Energy bands for the upper t$_{2g}$ `doublet' complex from the tight binding fit to
the full relativistic calculation of Na$_4$Ir$_3$O$_8$.  The horizontal line marks $E_F$.}
\label{fig8}
\end{figure}

\begin{figure}
\centerline{
\includegraphics[width=3.4in]{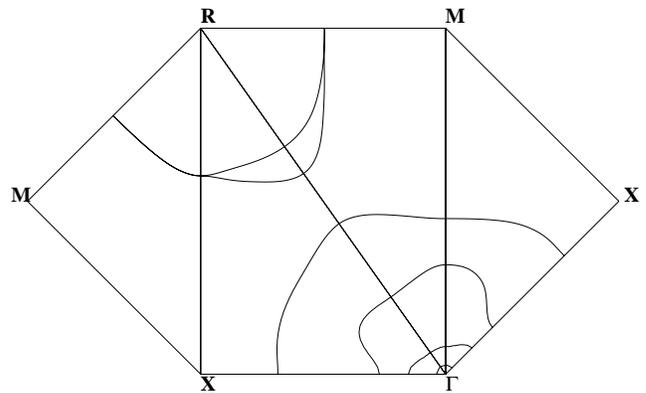}
}
\caption{Fermi surface from the tight binding fit to the full relativistic calculation of Na$_4$Ir$_3$O$_8$.}
\label{fig9}
\end{figure}

\begin{table}
\caption{Tight binding hopping parameters in eV.  The on-site energies
are $\epsilon_{O1}$ = -5.1913,
$\epsilon_{O2}$ = -4.0954, $\epsilon_{t_{2g}}$ = -1.7983, $\epsilon_{e_{g}}$ = 0.7987,
with the spin-orbit coupling $\lambda$ = 0.6386.
}
\begin{ruledtabular}
\begin{tabular}{lccc}
 & $\sigma$ & $\pi$ & $\delta$ \\
\colrule
Ir-O1 &  -2.5579  & 0.3186 & \\
Ir-O2 &  -2.1070  & 1.1817 & \\
Ir-Ir & -0.6372  & 0.0719  & 0.1545 \\
O1-O1 & 0.6156  & 0.0354 & \\
O2-O2 & 0.5463  & -0.2121 &\\
O1-O2 & -1.1327 &  0.0663 & \\
\end{tabular}
\end{ruledtabular}
\end{table}

\section{Exchange integrals}

\begin{table}
\caption{Exchange integrals in meV.  The spin Hamiltonian is given in Eq.~1.
Quoted are values where Ir site $m$ is along an (0,1,-1) direction relative to Ir site $n$.
$D$ are the Dzyaloshinski-Moriya, and
$\Gamma$ the anisotropic superexchange terms (in the second row for $\Gamma$,
the $jk$ refer to the parenthesis, $xy$, etc.).  In addition, the direct exchange is $J_d$ = 20.1, and
the isotropic superexchange term is $J_s$ = 12.9.  The last row, $J_{ii}$, is the total exchange for the
diagonal components ($J_d + J_s + \Gamma_{ii}$).
The assumed value of $U$ is 0.5 eV \cite{jjyu} (with $U_p$ set to zero).
}
\begin{ruledtabular}
\begin{tabular}{lccc}
$i~(jk)$ & $x~(xy)$ & $y~(xz) $ & $z~(yz)$ \\
\colrule
D$_i$  & 47.3  & 1.3 & -4.7 \\
$\Gamma_{ii}$ & 36.1  & -36.8  & -36.2 \\
$\Gamma_{jk}$ & 2.1  & -8.3  & -0.2 \\
$J_{ii}$ & 69.2  & -3.7  & -3.1 \\
\end{tabular}
\end{ruledtabular}
\end{table}

As we are in the strong spin-orbit coupling limit in the sense
of Chen and Balents,\cite{balents} we can use the expressions in their paper to estimate
the various exchange integrals (Table III).  We caution that these values are very
sensitive to the tight binding parameters, as well as the values of the Coulomb $U$.
The exchange Hamiltonian is of the form \cite{balents}
\begin{equation}
H_{ex} = (J_d + J_s) {\bf S}_n \cdot {\bf S}_m + {\bf D}^{nm} \cdot ({\bf S}_n \times {\bf S}_m)
+ {\bf S}_n \cdot {\bf \Gamma}^{nm} \cdot {\bf S}_m
\end{equation}
where $J_d$ is the direct exchange between iridium ions, $J_s$ the isotropic component
of the superexchange mediated by the oxygen ions, $D$ the Dzyaloshinski-Moriya
interaction, and $\Gamma$ the anisotropic part of the superexchange.

Remarkably, for the direct
Ir-Ir exchange, we find that even for the distorted lattice, the interaction is isotropic, and is given
by the simple formula $J_d = 2t^2/U$, where $t = \frac{1}{4}t_{dd}^{\sigma}+\frac{1}{3}t_{dd}^{\pi}
+\frac{5}{12}t_{dd}^{\delta}$.  The value we find is sizable, about 20 meV, which is not so
different from the experimental estimate of 28 meV.\cite{okamoto}  It would be larger,
except for the anomalously large value of $t_{dd}^{\delta}$.  In that context, we remark
that these hoppings are effective values, and compensate for the fact that we only include 18
tight binding parameters in our fits.  For instance, in the more complete analysis Mattheiss did for
the simpler IrO$_2$, $t_{dd}^{\delta}$ is only about 6\% of $t_{dd}^{\sigma}$.\cite{iro2}  We
have done a tight binding fit including the residual crystal field splittings (Fig.~3)
which did have a much reduced
value for $t_{dd}^{\delta}$ (with a larger value for $J_d$ of 37 meV), but the Fermi surface was 
degraded in this fit.

We now turn to the superexchange, which is responsible for all values of the exchange
Hamiltonian except
for $J_d$.  The values (Table III) are large and highly anisotropic
(we should caution, though, that these values 
will be further affected by the t$_{pp}$ hoppings \cite{shitade}).  This is unlike what would occur if the
local site symmetry of the Ir ions was purely cubic.\cite{podolsky}  The large values are not
only due to the distortion of the octahedra, but also because the superexchange between two
given Ir ions is mediated by an O1 and an O2 ion, each of which is in a different environment
(Fig.~10).
For instance, the non-zero values of $J_s$, $D$, and the off-diagonal components of $\Gamma$
are due to the distortion, with the magnitude strongly influenced by the difference between the O1
and O2 ions.
On the other hand, the reversed sign of the diagonal components
of $\Gamma$ relative to the cubic model of Chen and Balents \cite{balents} is due to the
the difference between O1 and O2, with the magnitude little affected by the distortion.

\begin{figure}
\centerline{
\includegraphics[width=2.4in]{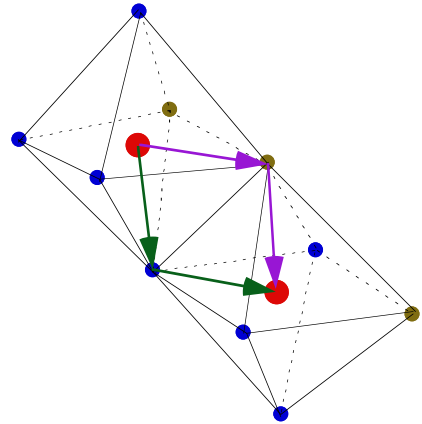}
}
\caption{(Color online) Superexchange pathway between two Ir ions as marked by the arrows.
The Ir ions are the large (red) spheres and the oxygen ions the small ones.  The lower left (green) 
pathway is via an O2 (blue) ion, the upper right (purple) pathway via an O1 (brown) ion.}
\label{fig10}
\end{figure}

There are only two ways we know to reduce these large values of the superexchange.
First, in the limit of large $U_p$
(Coulomb repulsion on the oxygen sites),
they would vanish.  Second, because the sodium sites about the oxygens are only partially occupied
in the real material, it is possible that the resulting randomness would dephase the various
superexchange pathways.  
Unless one of these mechanisms is effective, the total exchange is predicted
to be anisotropic.  This would cast doubts on theories for Na$_4$Ir$_3$O$_8$ based on an 
isotropic Heisenberg model.
Regardless, we believe that the tight
binding parameters we provide here will be useful for future modeling of this fascinating material.
In particular, given the resemblance of its electronic structure to that of cuprates,
one might wonder whether a doped version of this material would be a high
temperature superconductor.

\acknowledgments

Work at Argonne National Laboratory was supported by the U.S. DOE, Office of Science, under 
Contract  No.~DE-AC02-06CH11357.


\begin{thebibliography}{99}

\bibitem{leeNV} P. A. Lee, Science {\bf 321}, 1306 (2008).

\bibitem{phil73} P. W. Anderson, Mater. Res. Bull. {\bf 8}, 153 (1973).

\bibitem{okamoto} Y. Okamoto, M. Nohara, H. Aruga-Katori and H. Takagi,
Phys. Rev. Lett. {\bf 99}, 137207 (2007).

\bibitem{lawler} J. M. Hopkinson, S. V. Isakov, H.-Y. Kee and Y. B. Kim,
Phys. Rev. Lett. {\bf 99}, 037201 (2007);
M. J. Lawler, H.-Y. Kee, Y. B. Kim and A. Vishwanath,
Phys. Rev. Lett. {\bf 100}, 227201 (2008);
M. J. Lawler, A. Paramekanti, Y. B. Kim and L. Balents,
Phys. Rev. Lett. {\bf 101}, 197202 (2008);
Y. Zhou, P. A. Lee, T.-K. Ng and F.-C. Zhang,
Phys. Rev. Lett. {\bf 101}, 197201 (2008).

\bibitem{balents} G. Chen and L. Balents, Phys. Rev. B {\bf 78}, 094403 (2008).

\bibitem{pickett}
K.-W. Lee and W. E. Pickett, Phys. Rev. B {\bf 72}, 174505 (2005).

\bibitem{LMTO}
O. K. Andersen, Phys. Rev. B {\bf 12}, 3060 (1975).

\bibitem{HL}
L. Hedin and B. I. Lundqvist, J. Phys. C {\bf 4}, 2064 (1971).

\bibitem{SPLINE}
D. D. Koelling and J. H. Wood, J. Comp. Phys. {\bf 67}, 253 (1986).

\bibitem{LT}
G. Lehmann and M. Taut, Phys. Stat. Sol. (b) {\bf 54}, 469 (1972).

\bibitem{jjyu}
B. J. Kim, H. Jin, S. J. Moon, J.-Y. Kim, B.-G. Park, C. S. Leem, J. Yu, T. W. Noh, C. Kim, 
S.-J. Oh, J.-H. Park, V. Durairaj, G. Cao and E. Rotenberg,
Phys. Rev. Lett. {\bf 101}, 076402 (2008);
H. Jin, H. Jeong, T. Ozaki and J. Yu, Phys. Rev. B {\bf 80}, 075112 (2009).

\bibitem{phil87} P. W. Anderson, Science {\bf 235}, 1196 (1987).

\bibitem{SK}
J. C. Slater and G. F. Koster, Phys. Rev. {\bf 94}, 1498 (1954).

\bibitem{NR}
W. H. Press, B. P. Flannery, S. A. Teukolsky and W. T. Vetterling, {\it Numerical Recipes}
(Cambridge University Press, Cambridge, 1989).  The function being minimized is a sum
of the squares of the differences of the tight binding and band eigenvalues.

\bibitem{iro2}
L. F. Mattheiss, Phys. Rev. B {\bf 13}, 2433 (1976).

\bibitem{shitade}
A. Shitade, H. Katsura, J. Kunes, X.-L. Qi, S.-C. Zhang and N. Nagaosa,
Phys. Rev. Lett. {\bf 102}, 256403 (2009).

\bibitem{podolsky}
D. Podolsky and Y. B. Kim, arXiv:0909.4546.

\end{thebibliography}
\end{document}